\documentstyle[12pt,procsla,epsf]{article}
\def\lsim{\mathrel{\raise.2ex\hbox{$<$}\hskip-.8em\lower.9ex\hbox{$\sim$}}}
\def\gsim{\mathrel{\raise.2ex\hbox{$>$}\hskip-.8em\lower.9ex\hbox{$\sim$}}}
\def\mev{{\rm MeV}}

\def\km{{\rm km}}
\def\erg{{\rm erg}}
\def\mpc{{\rm Mpc}}
\def\m{{\rm m}}
\def\msec{{\rm msec}}

\def\cm{{\rm cm}}
\def\khz{{\rm kHz}}

  \catcode`\@=11
  \long\def\@makefntext#1{
  \protect\noindent \hbox to 3.2pt {\hskip-.9pt 
  $^{{\ninerm\@thefnmark}}$\hfil}#1\hfill}		%CAN BE USED
 
  \def\@makefnmark{\hbox to 0pt{$^{\@thefnmark}$\hss}}  %ORIGINAL
  	
  \def\ps@myheadings{\let\@mkboth\@gobbletwo
  \def\@oddhead{\hbox{}
  \rightmark\hfil\ninerm\thepage}  
  \def\@oddfoot{}\def\@evenhead{\ninerm\thepage\hfil
  \leftmark\hbox{}}\def\@evenfoot{}
  \def\sectionmark##1{}\def\subsectionmark##1{}}
 
\begin{document}
 
\font\fortssbx=cmssbx10 scaled \magstep2
\hbox to \hsize{
\includegraphics{/NextLibrary/TeX/tex/inputs/uwlogo.ps}
\hskip.5in \raise.1in\hbox{\fortssbx University of Wisconsin - Madison}
\hfill\vbox{\hbox{\bf MADPH-96-937}
            \hbox{March 1996}} }

\vspace{.5cm}

  \centerline{\normalsize\bf THE CASE FOR A KILOMETER-SCALE}
  \baselineskip=16pt
  \centerline{\normalsize\bf  HIGH ENERGY NEUTRINO DETECTOR: 1996\footnote{Talk presented by F. Halzen at the {\it Venitian Neutrino Conference}, February 1996.}}
  %\vfill
  %\vspace*{0.6cm}
  \centerline{\footnotesize F. Halzen}
  \baselineskip=13pt
  \centerline{\footnotesize\it Physics Department, University of Wisconsin,
   Madison, WI 53706}
  
  %\vfill
  \vspace*{0.5cm}
\begin{center}\small\bf Abstract\end{center}
{\footnotesize
The objective of neutrino astronomy, born with the identification of thermonuclear fusion in the sun and the particle processes controlling the fate of a nearby supernova, is to build instruments which reach throughout and far beyond our Galaxy and make measurements relevant to cosmology, astrophysics, cosmic-ray and particle physics. These telescopes will push astronomy to wavelengths smaller than $10^{-14}$~cm by mapping the sky in high-energy neutrinos instead of high-energy photons to which the Universe is partially opaque. While a variety of collaborations are pioneering complementary methods by building neutrino detectors with effective area in excess of 0.01~km$^2$, we show here that the science dictates 1~km$^2$, or a 1~km$^3$ instrumented volume, as the natural scale of a high-energy neutrino telescope. The construction of a high-energy neutrino telescope therefore requires a huge volume of very transparent, deeply buried material such as ocean water or ice, which acts as the medium for detecting the particles. We will speculate on its architecture. The field is immersed in technology in the domain of particle physics to which many of its research goals are intellectually connected. With several thousand optical modules the scope of constructing a kilometer-scale instrument is similar to that of experiments presently being commissioned such as the SNO neutrino observatory in Canada and the Superkamiokande experiment in Japan.
}
  
\normalsize\baselineskip=15pt
\setcounter{footnote}{0}
\renewcommand{\thefootnote}{\alph{footnote}}
 
\section{Introduction}

High-energy neutrino telescopes\cite{PR} are detectors whose architecture is optimized for achieving very large detection area rather than low energy threshold as was done for SNO and Superkamiokande designs. Two are partially deployed and operating: the lake Baikal detector and the Antarctic Muon and Neutrino Detector Array (AMANDA) at the South Pole. Two others, DUMAND and NESTOR, will be positioned at depths near 4 kilometers off the island of Hawaii and the coast of Greece, respectively. The science goals of these instruments are rich and truly interdisciplinary. We start by briefly enumerating the most important ones:

\begin{enumerate}

\item
{\bf Tomography of the universe}: it is important to realize that high-energy photons, unlike weakly interacting neutrinos, do not carry information on any cosmic sites shielded from our view by more than a few hundred grams of intervening matter. Because of diffuse backgrounds the Universe is partially opaque to photons with TeV energy and above. The TeV-neutrino sky could reveal objects with no counterpart in any wavelength of light. As was the case with radio telescopes, for instance, unexpected discoveries could be made. Forays into new wavelength regimes have historically led to the discovery of unanticipated phenomena. As is the case with new accelerators, observing the predictable will be slightly disappointing.

\item
{\bf Central engines of active galaxies}: although observations of PeV (10$^{15}$ eV) and EeV (10$^{18}$ eV) gamma rays are controversial, cosmic rays of such energies do exist and their origin is at present a mystery. The most energetic events exceed $10^{20}$~eV. Cosmic rays with energies up to some 10$^{14}$~eV are thought to be accelerated by shocks driven into the interstellar medium by supernova explosions. The Lorentz force on a particle near the speed of light in the galactic magnetic field (${\sim} 3 \mu$G) multiplied by the extent of a typical supernova shock (${\sim} 50$~pc) is only ${\sim}10^{17}$~eV. Our own Galaxy is too small, and its magnetic fields too weak, to accelerate particles to $10^{20}$~eV. This energy should require, for instance, a 100~$\mu$G field extending over thousands of light years. Such fields exist near the supermassive black holes which power active galactic nuclei (AGNs). This suggests the very exciting possibility that high-energy cosmic rays are produced in faraway galaxies and carry cosmological information --- on galaxy formation, for example.

The luminosity of AGNs often peaks at the highest energies, and their proton flux, propagated to Earth, can quantitatively reproduce the cosmic-ray spectrum above 10$^{18}$~eV\cite{bier}. Acceleration of particles is by shocks in the jets (or, possibly, in the accretion flow onto the supermassive black hole which powers the galaxy) which are a characteristic feature of these radio-loud, active galaxies. Inevitably, beams of gamma rays and neutrinos from the decay of pions appear along the jets. The pions are photoproduced by accelerated protons interacting with optical and UV photons in the galaxy which represent a target density of 10$^{14}$ photons per cm$^{3}$.

Observations of the emission of TeV (10$^{12}$ eV) photons from the giant elliptical galaxy Markarian 421\cite{punch} may represent confirming evidence of this scenario. Why Mrk 421? Although Mrk 421 is the closest of these AGNs, it is one of the weakest. The reason its TeV gamma rays are detected whereas those from other, more distant, but also more powerful, AGNs are not, must be that the TeV gamma rays suffer absorption in intergalactic space through the interaction with background infrared photons. The absorption is, however, minimal for Mrk 421 with a redshift $z$ as small as 0.03. In a study of nearby galaxies the Whipple instrument detected TeV emission from the blazar Mrk 501 with redshift $z=0.018$, a source which escaped the scrutiny of the Compton GRO observatory. All this strongly suggests that many AGNs may have significant, very-high-energy components, but that only Mrk 421 and 501 are close enough to be detected by gamma-ray telescopes. The opportunities for neutrino astronomy are wonderfully obvious. It is likely that neutrino telescopes will contribute to the further study of the high-energy astrophysics pioneered by space-based gamma-ray detectors, such as the study of gamma-ray bursts and the high-energy emission from quasars.

\item
{\bf Search for the accelerators of the highest-energy cosmic rays}: in heaven, as on Earth, high-energy neutrinos are produced in beam dumps which consist of a high-energy proton accelerator and a target. Gamma rays and neutrinos are generated in roughly equal numbers by the decay of neutral and charged pions produced in nuclear cascades in the beam dump. Neutrino telescopes can search for the neutrinos that accompany the acceleration and production of the
highest-energy cosmic rays, whether or not they are produced in AGN. It is, in general, important to realize that in efficient cosmic beam dumps with an abundant amount of target material, high-energy photons may be absorbed before escaping the source. Laboratory neutrino beams are an example. Therefore, the most spectacular neutrino sources may have no counterpart in high-energy gamma rays.

\item{\bf Neutrinos produced by interaction of cosmic rays with background photons}: by their very existence, high-energy cosmic rays guarantee the existence of sources of high-energy cosmic neutrinos. Cosmic rays represent a beam of known luminosity, with particles accelerated to energies in excess of $10^{20}$~eV. They produce pions in interactions with the Earth's atmosphere, the sun and moon, interstellar gas in our galaxy, and the cosmic photon background in our Universe. These interactions are guaranteed sources of calculable fluxes of photons and neutrinos\cite{PR}.

The study of extremely energetic, diffuse neutrinos produced in the interactions of the highest-energy, extra-galactic cosmic rays with the microwave background is of special interest. The magnitude and intensity of this cosmological neutrino flux are determined by the maximum injection energy of the ultra-high-energy cosmic rays and by the distribution of their sources. If the sources are relatively near, at distances of order tens of Mpc, and the maximum injection energy is not much greater than the highest observed cosmic-ray energy (few${}\times 10^{20}$~eV), the generated neutrino fluxes are small. If, however, the highest-energy cosmic rays are generated by many sources at large redshift, then a large fraction of their injection energy would be presently contained in gamma-ray and neutrino fluxes. The effect may be further amplified if the source luminosity were increasing with redshift $z$, i.e.\ if cosmic-ray sources were more active at large redshifts --- ``bright-phase models"\cite{berez}.

\item{\bf Study of neutrino oscillations by monitoring the atmospheric neutrino  
beam}: recent underground experiments have given tantalizing hints for neutrino oscillations in the mass range $\Delta m^2 \gsim 3\times 10^{-2}\rm\ eV^2$\cite{PR}. High-energy neutrino telescopes may be able to study and extend this mass range by measuring the zenith angle distribution of atmospheric neutrino-induced muons. For angles of arrival of atmospheric neutrinos ranging from vertically upward to downward, the neutrino path length (distance from its production to its interaction in the deep detector) ranges from the height of the atmosphere ($\sim$10~km) to the diameter of the Earth (${\sim}10^4$~km). With sufficient energy resolution it is possible to observe the oscillatory behavior of the flux over the oscillation length of several hundred kilometers suggested by the ``atmospheric neutrino anomaly". Only a mature and well-calibrated instrument can be expected to do this  
precision measurement.

Kilometer-scale detectors can determine neutrino flavor, as will be discussed in the next section.
The discovery of neutrino beams of cosmological origin would provide us with the opportunity to study neutrino oscillations over distances of billions of light years.  Also, possible observation of coincident gamma ray bursts (GRBs) of neutrinos and gamma rays can be used to make a measurement of the neutrino mass using a method well-advertised in connection with supernova 1987A. The mass is determined from the time delay $t_d$ between the neutrino and (massless) photon signals by simple relativistic kinematics, with $m_{\nu}=E_{\nu}\sqrt{2ct_d\over D}$. With $t_d$ possibly of order milliseconds, distances $D$ of billions of light years and energies $E_{\nu}$ similar to that of a supernova, neutrino observations from GRBs could improve the limit obtained from supernova 1987A by a factor $10^5 \sim 10^6$. The sensitivity of order $10^{-4} \sim10^{-5}$~eV is close to the range implied by the solar neutrino anomaly. The measurement would be greatly facilitated by the fact that, unlike for rare supernova events, repeated observations are possible. 

\item
{\bf Search for halo dark matter}: an ever-increasing body of evidence suggests that cold dark matter particles constitute the bulk of the matter in the Universe. Big-bang cosmology implies that these particles have interactions of order the weak scale, i.e.\ they are WIMPs\cite{Drees}. We know everything about these particles (except whether they really exist!). We know that their mass is of order of the weak boson mass; we know that they interact weakly. We also know their density and average velocity given that they constitute the dominant component of the density of our galactic halo as measured by rotation curves. WIMPs will annihilate into neutrinos with rates that are straightforward to estimate; {\it massive} WIMPs will annihilate into {\it high-energy} neutrinos which can be identified by neutrino telescopes. This technique has been recognized as a powerful tool to search for supersymmetry which predicts the existence of a stable, weakly interacting particle and, therefore, a very attractive dark matter candidate.  

\item
{\bf Gamma ray astronomy with neutrino telescopes}: the versatility of neutrino telescopes has been dramatically illustrated by the recent suggestion\cite{Stanev} to use neutrino detectors as gamma-ray telescopes. Underground detectors are designed to measure the directions of up-coming muons of neutrino origin. They can, of course, also observe down-going muons which originate in electromagnetic showers produced by gamma rays in the Earth's atmosphere. Although gamma-ray showers are muon-poor, it can be shown that they produce a sufficient number of muons to detect the sources observed by GeV and TeV gamma ray instruments. With a gamma-ray threshold higher by one hundred and a probability of muon production by the gammas of about $1\%$, even the shallower AMANDA and Lake Baikal detectors with a low muon threshold, have to overcome a $10^{-4}$ handicap. They can nevertheless match the detection efficiency of a GeV-photon satellite detector because their effective area is larger by a factor $10^4$ or more. 

Muons originate in TeV gamma showers whose existence has been established by air-Cherenkov telescopes.  They leave tracks in the detectors that can be adequately reconstructed by the Cherenkov technique and the direction of the parent photon can be inferred with degree accuracy. A multi-TeV air shower will produce a 100~GeV muon with a probability of order 1\%, sufficient to observe the brightest sources using relatively modest size detectors with effective area of order 1000~m$^2$ or more. Although muons from such sources compete with a large background of down-going cosmic-ray muons, they can be identified provided the detectors achieve sufficient angular resolution.
 
Unlike air Cherenkov telescopes, muon detectors cover a large fraction of the sky with a large duty cycle, e.g\ 100$\%$ for half of the sky in the case of the AMANDA detector with a South Pole location. The advantage is considerable in studying the emission from highly variable high-energy sources. For a detailed discussion of signals and cosmic ray background rejection; see Ref.~6.

\item
{\bf Burst detection of supernovae and GRBs}: high-energy neutrino telescopes deploy Optical Modules (OMs) in a clear medium which acts as the radiator of Cherenkov light from muon tracks or electromagnetic showers produced in neutrino interactions. They are primarily designed to exploit the long range of high-energy muons and have typical nominal threshold energies in the GeV-range, ostensibly too high to observe supernova neutrinos. Observation is nonetheless possible. For the 10~second duration of a supernova the interactions of  $\bar\nu_e$ with protons produces copious numbers of positrons with tens of MeVs of energy. These will yield excess signals in all OMs for the short duration of the burst. Such an observation, even if statistically weak in a single OM, will become significant for a sufficient number of OMs. The number of OMs required for monitoring galactic supernovae has been shown to be of order a few hundred, a number typical for the first-generation detectors under construction\cite{halzen}. In the ultra-clear South Pole ice for instance, the effective volume of a single OM is similar to that of the Kamioka detector, and a supernova near the center of the galaxy will yield over 20,000 events for a statistical significance in excess of $20\sigma$.

The same technique can be used to search for neutrinos from GRBs. The origin of gamma ray bursts (GRBs) is arguably astronomy's most outstanding puzzle. Contributing to its mystery is the failure to observe counterparts in any other wavelength of light. It should therefore be a high priority to establish whether GRBs emit most of their energy in neutrinos as expected in the (presently favored) cosmological models. We discuss this in more detail in the last section.

\item
{\bf Environmental studies}: in order to operate as a natural medium for particle detection the properties of  lake and deep ocean water and ice have to be studied with unprecedented precision. Initial studies of deep lake Baikal water and South Pole ice already have already resulted into totally unanticipated discoveries. AMANDA measurements revealed deep ice as the purest natural substance and the most transparent crystalline solid ever identified\cite{sweden}. 

\item
{\bf Earth tomography}: neutrino beams, including the atmospheric beam, may be used to do tomographic studies of the Earth's core. Near $10^{14}$~eV energy the absorption length of neutrinos becomes comparable to the diameter of the Earth. Neutrinos are attenuated along their travelling paths through the Earth and are therefore sensitive to its density profile. Studies of neutrino attenuation will lead to improved measurements of the Earth's core, which should complement seismic results that are often difficult to interpret\cite{Mann}.

\end{enumerate}

While the above science is very exciting, some of it is speculative. Neutrino detectors will however do a lot of ``bread and butter" physics. They will study the neutrinos and muons which are the byproducts of cosmic ray interactions in the atmosphere. There are other guaranteed sources in the neutrino sky. Cosmic rays interacting with the hydrogen in the galactic plane produce pions. MeV gamma rays from the decay of neutral pions have been observed by satellite detectors. Neutrinos from charged pions will delineate the galactic plane in the high-energy neutrino sky which we expect, however, to be dominated by extra-galactic sources.

\begin{figure}[t]
\centering
\hspace{0in}
\epsfxsize=4.2in\epsffile{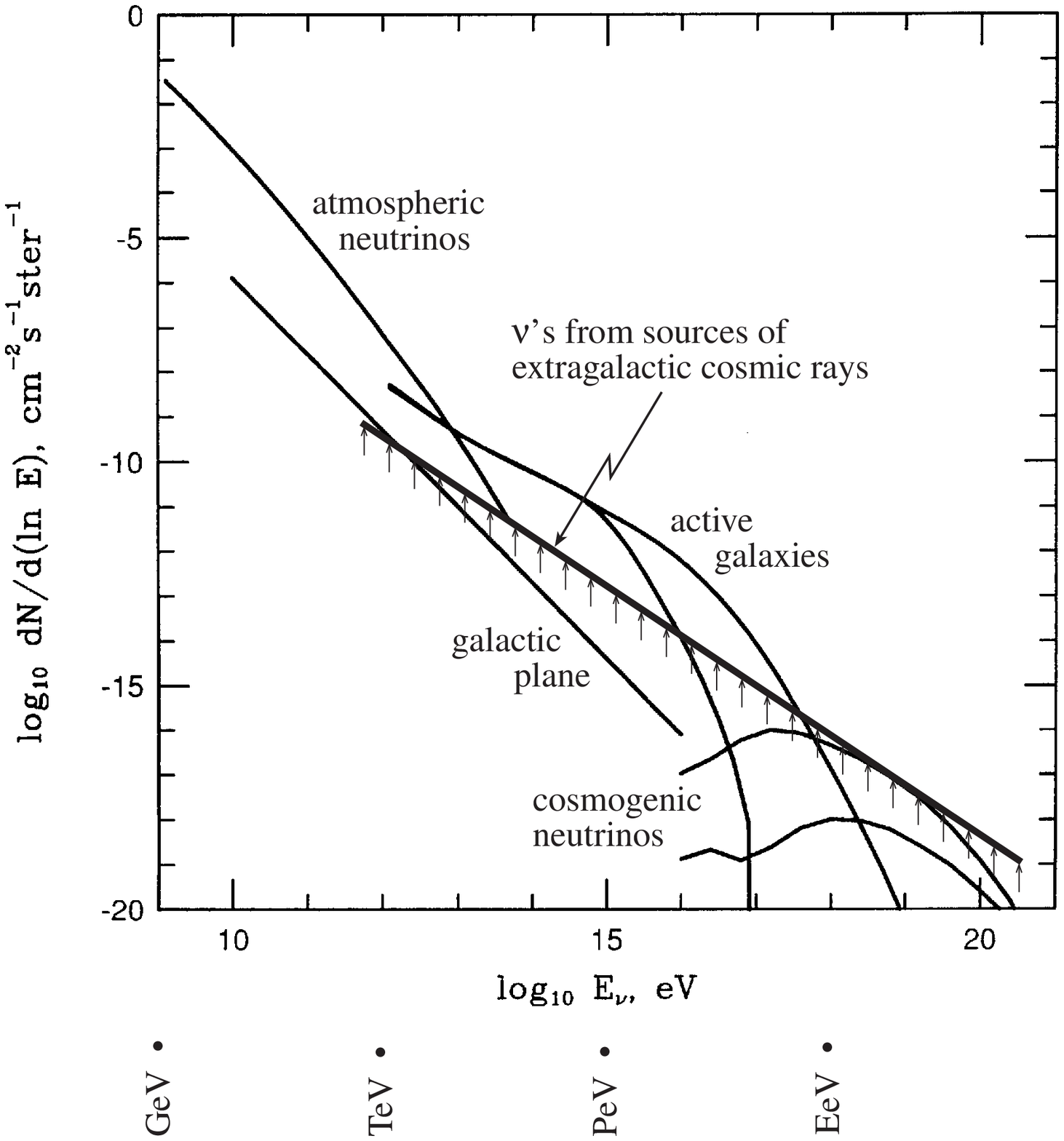}

\smallskip
{\footnotesize Fig.~1. Some cosmic sources of high-energy neutrinos.}
\end{figure}

The Milky Way may not be the prominent feature it is for visible light. The neutrino sky at GeV-energy and above is summarized in Fig.~1. Shown is the flux from the galactic plane as well as a range of estimates (from generous to conservative) for the diffuse fluxes of neutrinos from active galaxies and from the interaction of extra-galactic cosmic rays with cosmic photons. At PeV energies and above all sources dominate the background of atmospheric neutrinos. In order to deduce the effective area of an instrument required to study the fluxes in the figure, the detection efficiency must be included. At the highest energies this efficiency approaches unity and 1 event per km$^2$ per year corresponds to the naive estimate of $10^{-18}$ neutrinos per cm$^2$ second, barely sufficient to observe neutrinos produced by cosmic ray interactions with the cosmic microwave background. This is our first indication that a kilometer is the natural scale of a high-energy neutrino telescope, a fact first suggested 3 decades ago\cite{berez}. In the last section of this paper I will show how Nature has conspired to make a kilometer the natural scale for doing most of the science discussed above\cite{Snowmass}. The case can be made particularly succinct for AGNs, GRBs and the search for cold dark matter. We first give a conceptual description of the detectors.

\section{The Kilometer-Scale Neutrino Telescope: the ``Gedanken Experiment"}

In order to achieve the very large effective detection volumes required by the science, one optimizes the detector at high energies where: i) neutrino cross sections are large and the muon range is increased to several kilometers, ii) the angle between the muon and parent neutrino is less than $\sim$1 degree making astronomy possible, and iii) the atmospheric neutrino background is small. High-energy neutrino telescopes, just like the pioneering IMB and Kamiokande detectors, use phototubes to detect Cherenkov light from muons, but optimize their detector architecture to achieve large effective area at GeV-energy and above.  Inevitably the threshold is increased to $\sim$1~GeV or more above the MeV values characteristic for IMB and Kamiokande.

A ``gedanken" neutrino telescope is shown in Fig.~2a. It consists of a few thousand OMs viewing ocean, lake water, or ice. It is shielded from down-going fluxes by a kilometer or more. It ideally has a hybrid structure with a fine grid of OMs in the core surrounded by OMs on a coarse lattice; other architectures are possible\cite{Learned}. The core is instrumented over a kilometer cube, probably less. The OMs are spaced by distances of order 10~meter. The rest, of the volume of size a kilometer cube, possibly more, is coarsely instrumented with OMs spaced by 50$\sim$100~meters. Such instruments can be operated as a muon tracking device, a shower calorimeter and a burst detector. A cartoon of the various triggers is shown in Fig.~2b. We discuss them next.

\begin{figure}[h]
\centering
\hspace{0in}
\epsfxsize=4.25in\epsffile{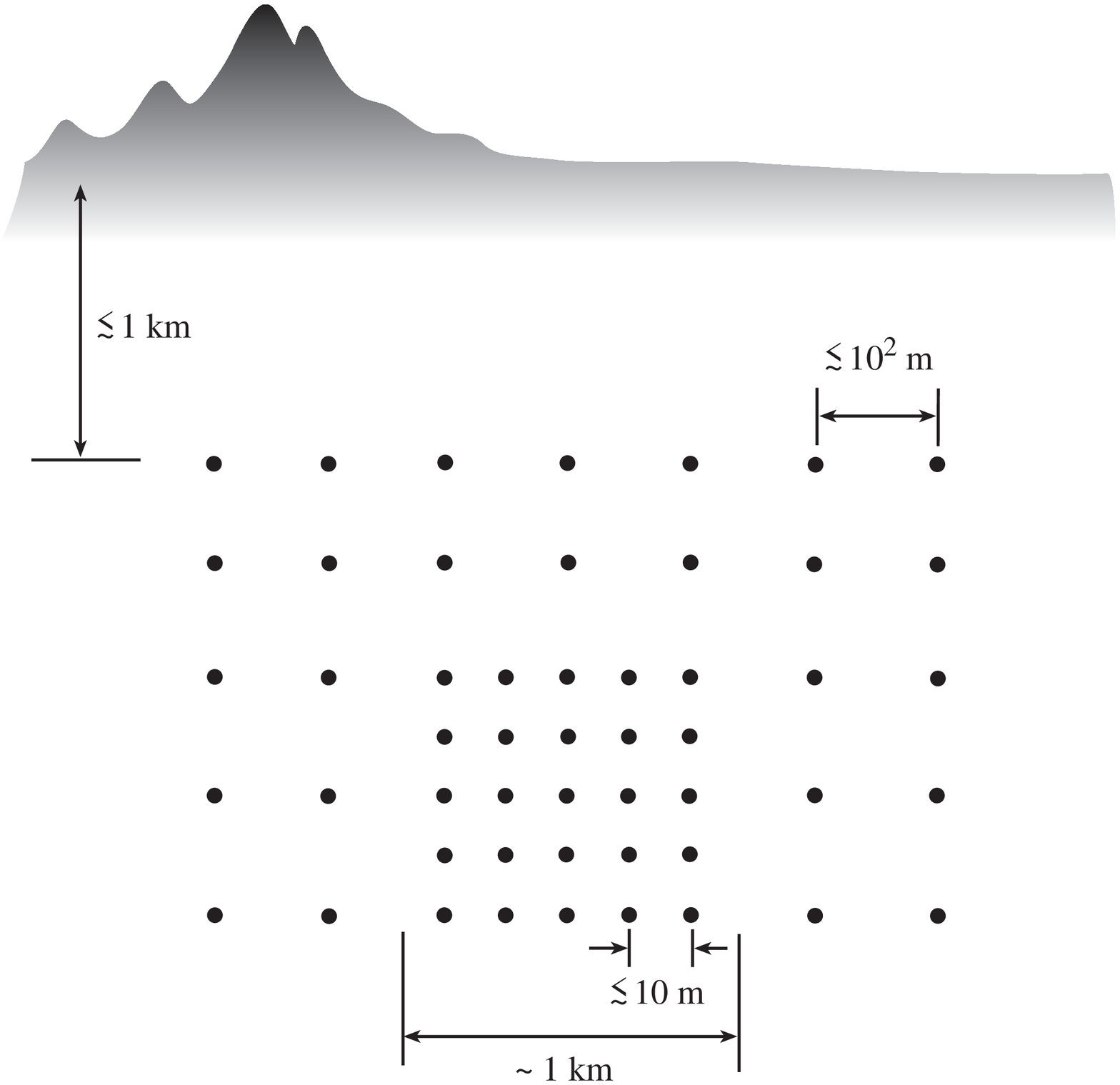}

\smallskip
{\footnotesize Fig.~2a. Hybrid neutrino telescope.}
\end{figure}

%% \begin{itemize}
\noindent$\bullet~$
In a Cherenkov detector the direction of the neutrino is inferred from the muon track which is measured by mapping the associated Cherenkov cone traveling through the detector. The arrival times and amplitudes of the Cherenkov photons, recorded by a grid of optical detectors, are used to reconstruct the direction of the radiating muon. The challenge is to record the muon direction with sufficient precision to unambiguously separate the much more numerous down-going cosmic-ray muons from the up-coming muons of neutrino origin. This task must be performed using a minimum number of optical modules, typically 10 or less. Critical parameters are the absorption and scattering length in the Cherenkov medium, the depth of the detector, which determines the level of the cosmic-ray muon background, and the noise rates in the optical modules which will sprinkle the muon trigger with false signals. Loosely speaking the absorption length determines the trigger volume of the detector, while the scattering length determines the distance over which timing can be maintained with sufficient accuracy. Sources of noise include radioactive decays such as decay of potassium-40 in water, bioluminescence and, inevitably, the dark current of the photomultiplier tube.

\begin{figure}[h]
\centering
\hspace{0in}
\epsfxsize=4.25in\epsffile{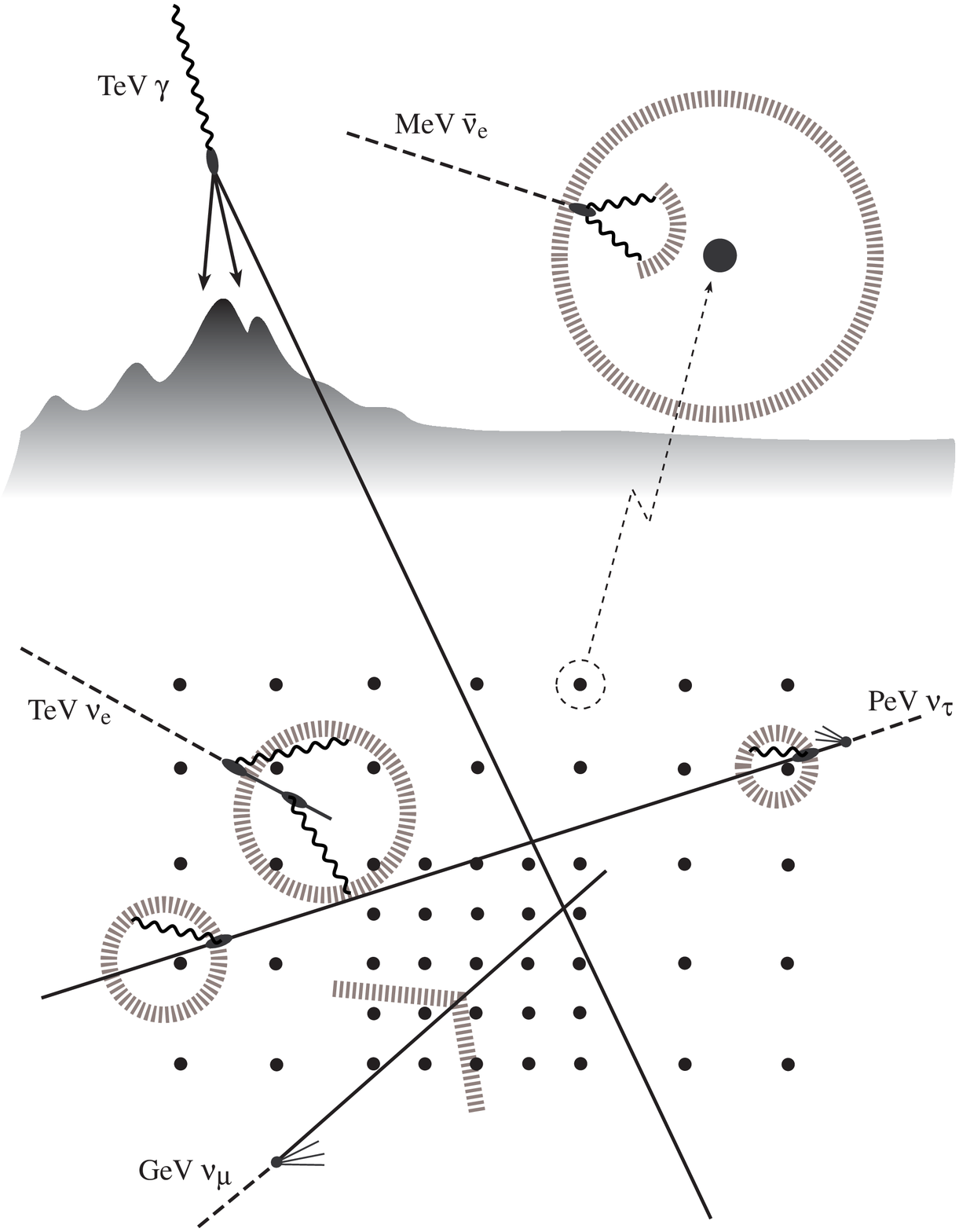}

\bigskip
\parbox{5.5in}{\footnotesize\baselineskip13pt Fig.~2b. Capabilities of a neutrino telescope: pictured are the detection of muons from gamma rays and neutrinos, showers initiated by high-energy electron neutrinos and by a low energy neutrino of supernova or gamma ray burst origin and, finally, a ``double bang" event from production and decay of a $\tau$ lepton.}
\end{figure}

\noindent
$\bullet~$
The grid of optical modules can also be used to map PeV electromagnetic showers initiated by electron neutrinos, e.g.\ PeV showers from the production of intermediate bosons in the interactions of cosmic electron neutrinos with atomic electrons in the detector. Mapping of showers can also be used to detect the bremsstrahlung of very-high-energy muons of neutrino origin as well as the production and decay of $\tau$-neutrinos. The ``double bang" event in Fig.~2b may represent the production and decay of a $\tau$ lepton by a TeV $\tau$-neutrino\cite{pakvasa}. The event can be differentiated from a muon track which loses energy catastrophically by the characteristic energy ratio of the 2 showers and by their spatial separation which reflects the  $\tau$ lifetime. 

Notice that {\it there is no atmospheric background for showering events once their energy exceeds 10~TeV}, although the precise value is model-dependent; see Fig.~1. Detection of neutrinos well above this energy would constitute the discovery of cosmic sources.

\noindent
$\bullet~$
The passage of a large flux of MeV neutrinos from a supernova or gamma ray burst will be detected as an excess of single counting rates in all individual optical modules of a neutrino telescope. The capabilities of neutrino telescopes to detect such bursts will be discussed in the next section.

%%\end{itemize}

It is revealing to note that the present activities of the AMANDA collaboration are dominated not only by muon reconstruction, but also by the search for TeV--PeV showers and the implementation of the supernova watch and GRB triggers\cite{sweden}. These topics are not even mentioned in a proposal written 4 years ago. This is exploratory science and surprises should be expected.  The instrumentation itself, frozen into the deep ice, will not become obsolete, and the electronics and logic located at the surface can be updated as new ideas arise. Whether in water or ice, the infrastructure should be viewed as a facility, like a telescope or an accelerator, whose missions can be updated following new science. Even architectures may be adjusted during construction, following progress (and hopefully discoveries) with partially deployed detectors. Unlike ice detectors, a water detector may be able to reposition its strings even after deployment.

Given a number of OMs, design choices typically fall between extremes: dense packing of the OMs in order to achieve good angular resolution and low threshold, or instrumenting the largest volume of ice in order to achieve large telescope area. Both are incorporated in the hybrid detector shown in Fig.~2a.

\smallskip\noindent
$\bullet~${\bf Dense-Pack Architecture}.
This approach, pioneered by the DUMAND\cite{wilkes} and Baikal\cite{spiering} experiments, achieves the lowest thresholds, perhaps GeV-energy, and is therefore ideal for WIMP or neutrino mass searches. Such detectors only reach large effective area at the higher energies by detecting muons produced far outside the instrumented volume. The range of TeV muons is indeed several kilometers. A well-known handicap of this approach is that the energy of the muon is only determined on a logarithmic scale. Energy is obtained in quantized 1,10,100,1000 TeV increments, e.g.\ by measuring the number of optical modules triggered by the muon. Further problems arise because of the confusion of an energetic muon with a bundle of low energy ones.

\smallskip\penalty-2000\noindent
$\bullet~${\bf Distributed Architecture}.
The alternative approach where large volumes are instrumented with widely spaced OMs looks very promising and is being actively investigated by the AMANDA collaboration\cite{sweden}. When the instrumented array dimensions approach 1 kilometer the sensitivity of a search for PeV neutrinos is about the same whether one detects cascades or muon tracks.  For smaller arrays this is not true since cascades are only observed in the relatively small instrumented volume, while one can detect muons which originate kilometers away. This benefit is reduced (and the challenge to reconstruct muons far outside the detector does not have to be met!) once  the array size is comparable to the muon range.  The power to search for AGN neutrinos is now similar for muon tracks or cascades. Obviously, the threshold has been raised because the muon has to lose energy catastrophically. The method emphasizes the search for the rare very-high-energy events, for instance those expected from active galaxies. 

The energy resolution of a coarse grid detector which identifies muons by their bremsstrahlung showers is expected to be much improved. The detector now has a linear response to energy. Once energy can be measured, the telescope is able to map the entire sky using only the most energetic events for which there is no atmospheric background. One is no longer limited to up-coming events.

Does a coarse grid preclude the observation of point sources?  Probably not.  Track reconstruction is much easier once the origin of one (or more!) cascades reveals a point (points) on the track. With one known vertex, only 2 out of 5 parameters (2 angles and 3 coordinates) are to be fitted. Muon tracks which radiate more than once are reconstructed with superb accuracy as two, or more, points on the track are pinpointed.

The discovery of absorption lengths of several hundred meters in ice seduces one to construct a cheap kilometer-scale detector with relatively high threshold, probably not much less than 1~TeV. The approach is reminiscent of proposals to detect acoustic or radiowave signals produced by PeV neutrinos or muons, although the threshold is lower  by several orders of magnitude. These methods were developed to exploit the large absorption length of acoustic and giga-Hertz radiowaves in water or ice, allowing the deployment of detector elements on a grid with large spacings. In the case of ice, Cherenkov light shares this property. From all other points of view light has significant advantages: PMTs represent a cheap and well-understood technology, the ambient backgrounds are understood and the threshold of the detector is lower by one or, most likely, several orders of magnitude. Monte Carlo simulation of the deep bubble-free ice suggest that a 10~TeV shower penetrates 8 attenuation lengths, or about 500~meters. Detection of the diffuse light in a kilometer-scale instrument will allow calorimetry with a resolution of perhaps 25\%. (Also from the experimental point of view a kilometer is the natural scale of a high-energy neutrino detector!) Even after 200~meters the first photons reaching the OM are not scattered (there are roughly $10^3$ photons left, not all of them have undergone scattering after 6 scattering lengths). The Cherenkov cone can be adequately mapped in a detector with relatively large string spacings, perhaps more than 50~meters.

 As illustrated by the previous discussion, commissioning a kilometer-scale detector may in some ways be easier than building a prototype which must be able to do exquisite track reconstruction in order to expand its trigger volume far outside the instrumented volume. There are however new challenges. The duration of triggered events increases with the physical size of the detector and so does, inevitably, the number of noise hits confusing trigger reconstruction. On a kilometer scale this becomes a problem, especially when using large OMs\cite{Okada}. In sterile ice the challenge is easier to meet though it is not to be ignored.

\section{Why a Kilometer: 3 Simple Examples}

``Back of the envelope" calculations are sufficient to associate the kilometer scale with high-energy neutrino astronomy. We will work through three simple examples covering the search for the accelerators of the highest-energy cosmic rays, the search for cold dark matter particles or WIMPs and, finally, the search for the cosmic sources of GRBs. To a particle physicist these represent the most fascinating puzzles in astronomy.

\smallskip\noindent
$\bullet~${\bf Neutrinos associated with the production and acceleration of the highest-energy cosmic rays}

A simple estimate of the AGN neutrino flux can be made by assuming that a neutrino is produced for every accelerated proton. This balance is easy to understand once one realizes that in astrophysical beam dumps the accelerator and production target form a symbiotic system. Although larger target mass may produce more neutrinos, it also decelerates the protons producing them. Equal neutrino and proton luminosities are therefore typical for the astrophysical beam dumps considered\cite{PR} and implies that:
\begin{equation}
4\pi \int dE (E \, dN_\nu/dE) \sim L_{CR} \sim 10^{-9}\rm\ TeV\  
cm^{-2}\ s^{-1} \;,				
\end{equation}
The luminosity $L_{CR}$ has been conservatively estimated from cosmic ray data by assuming that only the highest-energy component of the cosmic-ray flux above 10$^{18}$~eV is of AGN origin. These particles, with energies beyond the ``ankle" in the spectrum, are almost certainly extra-galactic and are observed with a $E^{-2.71}$ power spectrum. Assuming an $E^{-2}$ neutrino spectrum, the equality of cosmic-ray and neutrino luminosities implies:
\begin{equation}
E{dN_\nu\over dE} = {1\over4\pi} {10^{-10}\over E\,(\rm TeV)} \, \rm  
cm^{-2}\ s^{-1}\ sr^{-1} \;. \label{eq:flux}
\end{equation}
Roughly the same result is obtained by assuming equal numbers of neutrinos and protons rather than equal luminosities. The flux of Eq.~(\ref{eq:flux}) is at the low end of the range of fluxes predicted in models where acceleration is in shocks in the jet\cite{bier} and accretion disc\cite{szabpro,stecker}; see Fig.~1. It is clear that our estimate is rather conservative because the proton flux reaching Earth has not been corrected for absorption in ambient matter in the source and in the interstellar medium.

The probability to detect a TeV neutrino is roughly $10^{-6}$\cite{PR}. A neutrino detector with 10$^6$~m$^2$ effective area is therefore required for observing 100 upcoming muons per year, or maybe 10 from a nearby source. Model predictions often exceed this estimate by several orders of magnitude.

\smallskip\noindent
$\bullet~${\bf WIMPS from the Sun}

Standard cosmology dictates the properties of WIMPs.  Their mass is of order of the weak boson mass with
\begin{equation}
       \mbox{tens of GeV} < m_{\chi} < \rm several\ TeV \,. \label{GT}
\end{equation}
Lower masses are excluded by accelerator and (in)direct searches with existing detectors, while masses beyond several TeV are excluded by cosmological considerations. Their density and average velocity in our Galaxy is also known given the assumption that they constitute the dominant component of the density of our galactic halo as measured by rotation curves. 

Neutrino telescopes can infer the existence of WIMPs from observation of their annihilation into neutrinos. {\bf Massive} WIMPs will annihilate into {\bf high-energy} neutrinos which can be detected in high-energy neutrino telescopes. Detection is greatly facilitated by the fact that the sun represents a dense and nearby source of accumulated cold dark matter particles\cite{Drees}. Galactic WIMPs, scattering off nuclei in the sun, lose energy. They may fall below escape velocity and be gravitationally trapped. Trapped WIMPs eventually come to equilibrium temperature and accumulate near the center of the sun. While the WIMP density builds up, their annihilation rate into lighter particles increases until equilibrium is achieved where the annihilation rate equals half of the capture rate. The sun has thus become a reservoir of WIMPs which  annihilate mostly into heavy quarks and, for the heavier WIMPs, into weak bosons. The leptonic decays of the heavy quark and weak boson annihilation products turn the sun into a source of high-energy neutrinos with energies in the GeV to TeV range, rather than in the keV to MeV range typical for neutrinos from thermonuclear burning.

The reach of the detector is determined by the flux of solar neutrinos of WIMP origin which is determined by the mass of the WIMPs and by their elastic cross section on nucleons. In standard cosmology WIMP capture and annihilation interactions are weak and dimensional analysis is sufficient to compute the neutrino flux from the observed WIMP density in our galactic halo. The assumptions are

\begin{enumerate}

\item
that WIMPs represent the major fraction of the measured halo density, i.e.
\begin{equation}
\phi_\chi = n_\chi v_\chi = {0.4\over m_\chi} \, {\rm {GeV\over cm^3} \ 
3\times10^6 {cm\over s} } = {1.2\times10^7\over m_{\chi}(\rm in\ GeV)} \,\rm
cm^{-2} s^{-1} \;,
\label{phi chi}
\end{equation}

\item
that the WIMP-nucleon interaction cross section is given by dimensional analysis 
\begin{equation}
\sigma(\chi N) = \left(G_F m_N^2\right)^2 {1\over m_W^2} 
= 6\times10^{-42}\rm\,cm^2 \;,
\label{sigma chi N}
\end{equation}

\item
that WIMPs annihilate 10\% of the time into neutrinos (this is just the leptonic branching ratio of the final state particles in the dominant annihilation channels $\chi\bar\chi \to W^+W^-$ or $Q\bar Q$, where $Q$ is a heavy quark).

\end{enumerate}

Clearly the cross section for the interaction of WIMPs with matter is uncertain. Arguments can be invoked to raise or decrease it. Important points are that i) our choice represents a typical intermediate value for current models, ii) our result for event rates scales linearly in the cross section and can be easily reinterpreted.

The calculation is now straightforward and can be found in Ref.~18. The result is shown in Fig.~3. A kilometer-size detector probes WIMP masses up to the TeV-range, beyond which they are excluded by cosmological considerations. Especially for heavier WIMPs the indirect technique is powerful because underground high-energy neutrino detectors have been optimized to be sensitive in the energy region where the neutrino interaction cross section and the range of the muon are large. 

\begin{figure}[h]
\centering
\hspace{0in}
\epsfxsize=4.25in\epsffile{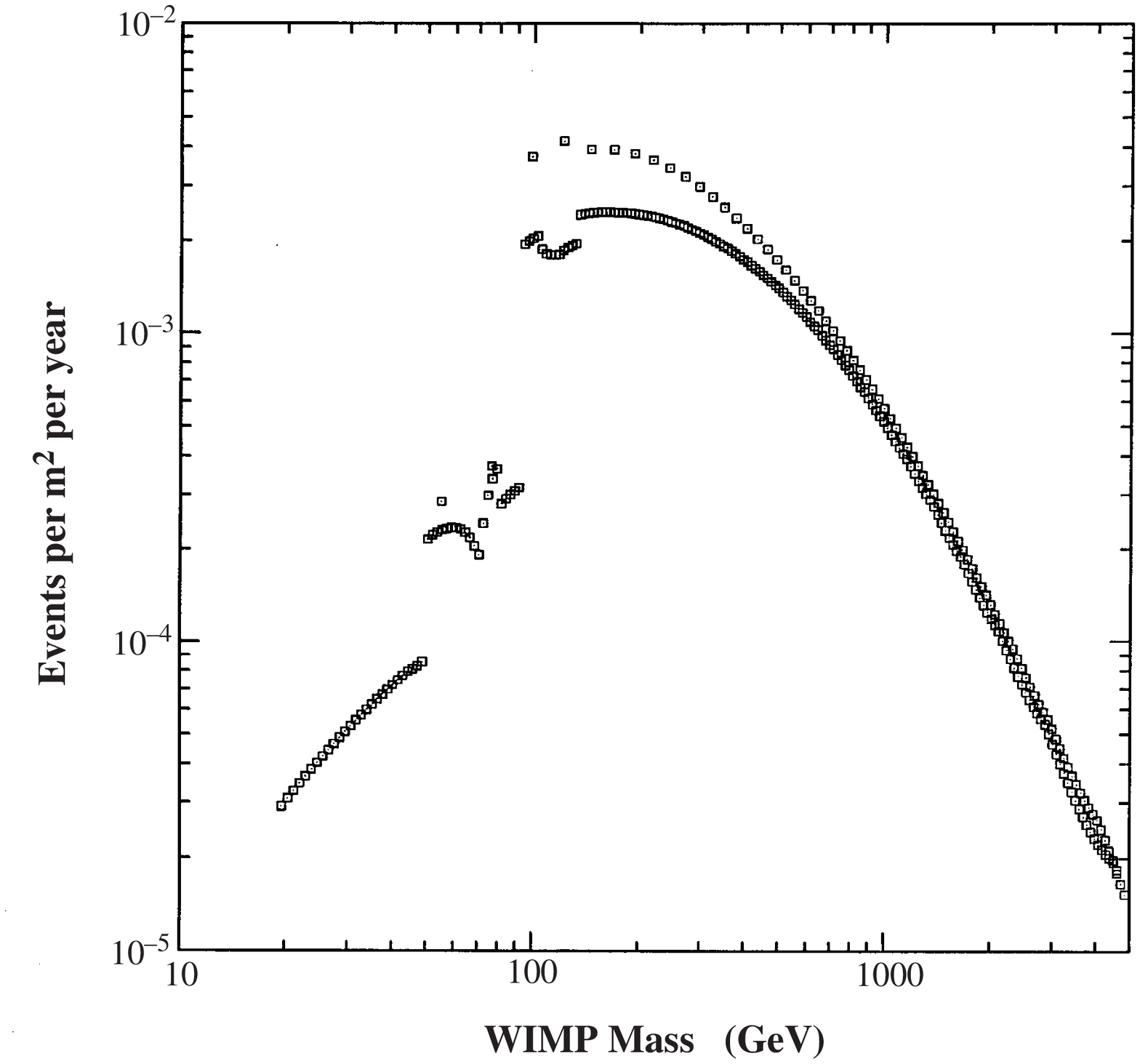}

\medskip
\parbox{5.5in}{\footnotesize Fig.~3. Flux of high-energy neutrinos produced by WIMP annihilation in the sun. The structure in the result is associated with particle thresholds and not important for present considerations.}
\end{figure}

For high-energy neutrinos the muon and neutrino are aligned, with good angular resolution, along a direction pointing back to the sun. The number of background events of atmospheric neutrino origin in the pixel containing the signal will be small. The angular spread of secondary muons from neutrinos coming from the direction of the sun is well described by the relation\cite{PR} $\sim 1.2^\circ \Big/ \sqrt{E_\mu(\rm TeV)}$. The muon energies and therefore, indirectly, the WIMP mass may be inferred from the angular spread of
the~signal.

\smallskip\noindent
$\bullet~${\bf Observing gamma ray bursts\cite{greg}}

The origin of gamma ray bursts (GRBs) is arguably astronomy's most outstanding puzzle. Contributing to its mystery is the failure to observe counterparts in any other wavelength of light. It should therefore be a high priority to establish whether GRBs emit most of their energy in neutrinos\cite{pacz,plaga,learned} as expected in the, presently favored, cosmological models. 

We will frame the problem in the context of the fireball models\cite{piran}. Although the details can be complex, the overall idea is that a large amount of energy is released in a compact region of radius $R\simeq 10^2$\,km$\simeq c \Delta t$. The shortest time-scales, with $\Delta t$ of order milliseconds, determine the size of the initial fireball\cite{piran}. 
Only neutrinos escape because the fireball is opaque to photons. It is indeed straightforward to show that the optical depth of the fireball is of order $10^{13}$ because a significant fraction of the photons are above pair production threshold and produce electrons.
 It is then theorized that a relativistic shock, with $\gamma \simeq 10^2$ or more, expands into the interstellar medium and photons escape only when the optical depth of the shock has been sufficiently reduced. The properties of the relativistic shock are a matter of speculation. They fortunately do not affect the predictions for neutrino emission.

For a gamma ray fluency $F=10^{-9}\rm\,J\,m^{-2}$ and a distance $z=1$ for typical bursts, the energy required is
\begin{equation}
E_\gamma = 2\times 10^{51}{\rm\, erg} \left(D\over 4000{\rm\; Mpc}\right)^2 
\left(F\over 10^{-9}\rm\,J\>m^{-2}\right) \,,  \label{eq:E_gamma}
\end{equation}
using $E_\gamma = 4\pi D^2 F$. The temperature $T_\gamma$ is obtained from the energy density
\begin{equation}
\rho = {E_\gamma\over V} = {1\over2} h a T^4 \,,
\end{equation}
where $h$ represents the degrees of freedom ($h_\gamma = 2$ and $h_\nu = 
2\cdot3\cdot{7\over8}$ for 3 species of neutrinos and antineutrinos), $V$ the 
volume corresponding to radius $R$ and $a = 7.6 \times 10^{-16}\rm\; J\; m^{-3} 
\,K^{-4}$. We find that
\begin{equation}
T_\gamma = 8\;\mev \left(E_{\gamma}\over 2\times10^{51}\,\erg\right)^{1/4} 
\left(100\;\km \over R\right)^{3/4} \,. \label{eq:T_gamma}
\end{equation}
For neutrinos
\begin{equation}
T_\nu = \left(E_\nu/E_\gamma\over h_\nu/h_\gamma\right)^{1/4} T_\gamma \,.
\label{eq:T_nu}
\end{equation}
For a merger of neutron stars, for instance, the release of a solar mass of energy of $2 \times 10^{53}$~erg implies a total energy emitted in neutrinos $\sim10^2E_\gamma$. The $\gamma$'s are most likely produced by bremsstrahlung off electrons produced by $\nu\bar\nu$ annihilation. The actual predictions for the energy and time structure of the photon signal depend on the details of the shock which carries them outside the opaque fireball region of size $R$. The data suggest that the structure of these shocks is complex. Neutrinos, on the 
contrary, promptly escape and carry direct information on the original explosion. From (\ref{eq:T_gamma}),(\ref{eq:T_nu}) we obtain $T_\nu\simeq2.5 T_\gamma\simeq20\;\mev$. Using this and a total neutrino energy in the fireball of $10^2E_\gamma$ we obtain the average energy of a neutrino of GRB origin:
\begin{eqnarray}
E_{\nu} &=& 3.15\ T_{\nu} \ =\ 65\;\mev \left(E_\gamma\over 
2\times10^{51}\,\erg\right)^{1/4} \left(100\;\km\over R\right)^{3/4} \,, 
\label{eq:E_nu_obs}\\
\Delta t_{\rm obs} &=& 0.3\;\msec \left(R\over 100\;\km\right) \,.
\end{eqnarray}
The neutrino fluency is obtained from $E_{\nu\rm\,tot} / (4\pi D^2)$
\begin{equation}
N_\nu = 10^4\,\m^{-2} \left(E_{\nu\rm\,tot}\over2\times10^{53}\,\erg\right) 
\left(65\;\mev\over E_\nu\right) \left(4000\;\mpc\over D\right)^2
\end{equation}
or more than $10^{57}\ \nu$'s at the source. We emphasize that this prediction is 
rather model-independent because it just relies on the fact that a solar mass 
of energy is released in a volume of 100~kilometer radius. This radius is determined by the observed duration of the bursts.

Although the $\sim100\,\mev$ neutrinos are below the muon threshold of high-energy 
neutrino telescopes, the $\bar\nu_e$ will initiate electromagnetic showers by the reaction  ($\bar\nu_e + p \to n + e^+)$ which will be counted by a supernova type trigger. Detailed 
simulations\cite{halzen} of the supernova signal in ice have shown that a 20~cm photomultiplier has a seeing radius $d\simeq7.5$~m for 20~MeV positrons, the average detected particle energy for a supernova burst. The seeing volume increases linearly with the energy of the positron. The number of events per PMT is given by
\begin{equation}
\#{\rm N_{\nu\,\rm obs}} \simeq N_{\nu} (\pi d^2) \left(d\over \lambda_{\rm 
int} \right) \,. \label{eq:events}
\end{equation}
The last factor estimates the probability that the $\bar\nu_e$ produces a 
positron within view of the PMT. Here the neutrino interaction length is given by
\begin{equation}
\lambda_{\rm int}^{-1} = {2\over18} A \rho \sigma_0 E_{\nu}^2
\end{equation}
for a $\bar\nu_e$ interaction cross section on protons:
\begin{equation}
\sigma_0 = 7.5\times10^{-40}\rm\,m^2\,MeV^{-2} \,.
\end{equation}
$A$ is Avogadro's number and $\rho$ the density of the detector medium. The event rate for GRBs is given by (\ref{eq:events}) with $d = 7.5\;\m\left(65\;\mev\over20\;\mev\right)^{1/3}$, making explicit the linear dependence of the volume on energy. Here $65\,\mev$ is the positron energy which is roughly equal to the neutrino energy given by Eq.~(\ref{eq:E_nu_obs}).

Can this signal be detected by simple counting of PMT hits? Signal $S$, noise $N$ and $S/\sqrt N$ are, for an average burst, given by
\begin{eqnarray}
S &=& 2 \times10^{-3}{\rm\;events} \left(N_{\nu\rm obs}\over 5 \times 10^{-6}\right) 
\left(D_{\rm PMT} \over 20\;\cm\right)^2 \left(N_{\rm PMT}\over 200\right) 
\,,\\
N &=& 60{\rm\;events} \left(\Delta t\over 0.3\;\msec \right) \left(N_{\rm 
back}\over 1 \khz\right) \left(N_{\rm PMT}\over 200\right) \,,\\
S/\sqrt N &=&3 \times 10^{-4} \left(N_{\rm back}\over 1\;\khz\right)^{-1/2} 
\left(D_{\rm PMT}\over 20\;\cm\right)^2 \left(N_{\rm PMT}\over200\right)^{1/2} 
\,.
\end{eqnarray}
AMANDA has been chosen for reference with roughly 200 PMTs in the trigger with a diameter $D_{\rm PMT}$ of 20~cm and a background counting rate of roughly 1~kHz. With such low 
rates in millisecond times, observation obviously requires a dedicated trigger.

\looseness=-1
The event rate for an {\it average} burst is predicted to be low. We
will argue nevertheless that observation is possible and clearly guaranteed
for kilometer-scale detector with several thousand PMTs. First, the parameters
entering the calculation are uncertain. The event rate increases with neutrino
energy as $E_{\nu}^3$ because of the increase of the PMT seeing distance $d$
and the neutrino interaction cross section $\sigma_0$. With increased energy
the average burst may become observable. Individual burst can yield orders of
magnitude higher neutrino rates because of intrinsically
higher luminosity and/or smaller than average distance to earth.
For example, a burst
10 times closer than average (which occurs every few years in cosmological models) and 10 times more energetic is observable with a
significance of well over 10~$\sigma$ in the existing AMANDA detector. Given
the uncertainties in the model and its parameters as well as the chaotic nature
of the phenomenon (there is no such thing as an average GRB), this event
represents a plausible possibility. Events at the $4 \sigma$ level can be expected to occur yearly.
  
\looseness=-1
As demonstrated by the  observations, the structure of the shock producing the gamma rays is complex. The interaction of multiple shocks can also produce neutrinos on other time-scales and with different, sometimes much higher, energies\cite{pacz}. It has been theorized\cite{katz} that when the relativistic shock runs into proton clouds in the interstellar medium, pions are produced which are the parents of the delayed GeV-energy gamma rays observed in some events. The charged pions produced in the same interactions will, inevitably, be a source of high-energy neutrinos. It is in particular unlikely that cosmic string models can escape the scrutiny of our search because they predict a fluency in neutrinos which exceeds that for photons by a factor of order $10^8$ or more.

This concludes our third argument for kilometer size telescopes with several thousand optical modules. There are many more\cite{PR,Snowmass}.

\newpage

\nonumsection{Acknowledgements}

I thank my AMANDA collaborators, especially Steve Barwick, for discussions. Sandip Pakvasa, Christian Spiering and Gaurang Yodh made constructive suggestions on the manuscript.
This work was supported in part by the University of Wisconsin
Research Committee with funds granted by the Wisconsin Alumni Research
Foundation, and in part by the U.S.~Department of Energy under Contract
No.~DE-AC02-76ER00881.


\begin{thebibliography}{999}
%
\let\sl=\it %% too many different fonts otherwise, looks messy
\frenchspacing

\bibitem{PR}
T.~K.~Gaisser, F.~Halzen and T.~Stanev, {\sl Physics Reports} {\bf 258}, 173 (1995)

\bibitem{bier}
K. Mannheim and P.L. Biermann, {\it Astron. Astrophys.} {\bf 22}, 211 (1989); K.~Mannheim and P.L. Biermann, {\it Astron. Astrophys.} {\bf 253}, L21 (1992); K.~Mannheim, {\it Astron.\ Astrophys.} {\bf 269}, 67 (1993).

\bibitem{punch}
M.~Punch {\em et al.}, {\sl Nature} {\bf 358}, 477--478 (1992).

\bibitem{berez}
See e.g., V.S.~Berezinsky {\em et al.}, {\sl Proc.\ of the Astrophysics of Cosmic Rays}, (Elsevier, New York, 1991).


\bibitem{Drees}
For recent reviews see e.g., M.~Drees and M.~M.~Nojiri,
 Phys.\ Rev.\ {\bf D47}, 376 (1993); G.~Jungman, M.~Kamionkowski, and K.~Griest, {\it Supersymmetric Dark Matter}, Physics Reports, to be published.

\bibitem{Stanev}
F. Halzen and T. Stanev, {\it Gamma ray astronomy with underground detectors}, University of Wisconsin-Madison preprint MADPH-95-901 [hep-ph/9507362].

\bibitem{halzen}
F. Halzen, J.E. Jacobsen, and E. Zas, Phys. Rev. {\bf D49}, 1758 (1994) and
University of Wisconsin-Madison preprint MADPH-95-888 [astro-ph/9512080]
(1995).


\bibitem{sweden}
AMANDA collaboration, Science {\bf 267}, 1174 (1995); J. Glaciology (in press);  
{\it Proceedings of the 24th International Cosmic Ray Conference}, Rome, 1995.

\bibitem{Mann}
A. K. Mann, University of Pennsylvania Preprint 0232E (1996).

\bibitem{Snowmass}
F. Halzen, {\it The case for a kilometer-scale neutrino detector}, in Nuclear and Particle Astrophysics and Cosmology, Proceedings of Snowmass~94, R.~Kolb and R.~Peccei, eds.

\bibitem{Learned}
J.~G.~Learned and A.~Roberts, {\it Proceedings of the 23$^{rd}$ International Cosmic Ray Conference}, Calgary, Canada  (1993);  F.~Halzen and J.G.~Learned, {\it Proc.\ of the Fifth
International Symposium on Neutrino Telescopes}, Venice (1993),
ed.\ by  M.~Baldo-Ceolin.


\bibitem{pakvasa}
J. G. Learned and S. Pakvasa, Astroparticle Phys. Journal {\bf 3}, 267 (1995).

\bibitem{wilkes}
R.J. Wilkes, in Proceedings of the 22nd Annual SLAC Summer Institute on Particle Physics, Stanford CA (1994).

\bibitem{spiering}
C. Spiering, Baksan School, Baksan (1995).

\bibitem{Okada}
A. Okada, talk presented at the LBL Meeting on Physics Issues for km$^3$ Neutrino Astronomy, Berkeley, CA (1994).

\bibitem{szabpro}
A.P. Szabo and R.J. Protheroe, in {\it Proc.\ of the High-Energy Neutrino Astrophysics Workshop}, Univ.\ of Hawaii, March 1992, eds.\ V.J. Stenger, J.G. Learned, 
S.~Pakvasa and X. Tata (World Scientific, Singapore).

\bibitem{stecker}
F.W.~Stecker, C.~Done, M.H.~Salamon and P.~Sommers, {\it Phys.\ Rev.\ Lett.} {\bf 66}, 2697 (1991) and {\bf 69}, 2738(E) (1992).

\bibitem{iowa}
F. Halzen, in {\it Proceedings of the International Symposium on Particle Theory and Phenomenology}, Iowa State University, Ames (1995)

\bibitem{greg}
F. Halzen and G. Jaczko, Phys. Rev., in press.

\bibitem{pacz}
B. Paczy\'nski and G. Xu, Astrophys. Journal {\bf 427}, 708 (1994).

\bibitem{plaga} R. Plaga, Astrophys. Journal {\bf 424}, L0 (1994).

\bibitem{learned}
J. Learned, S. Pakvasa, W.A. Simmons and T. Weiler, VAND-TH-94-20
[hep-ph/9411432] (1994).

\bibitem{piran}
For a review, see T. Piran, astro-ph/9507114 (1995).

\bibitem{katz}
J.I. Katz, Astrophys. Journal {\bf 432}, L27 (1994)





\end{thebibliography}
\end{document}